\documentclass{article}
\usepackage{authblk}
\title{Imágenes de Resonancia Magnética con Contraste en el Cáncer de Mama}
\author[1]{Virginia del Campo}
\author[2]{Iker Malaina}
\affil[1]{Hospital de Urduliz, OSI Uribe, Osakidetza\\
Goieta 32, 48610 Urduliz, Bizkaia, Spain}
\affil[2]{Departamento de Matemáticas, FCT, UPV/EHU\\
Sarriena sn, 48940 Leioa, Bizkaia, Spain}

\usepackage{amssymb}
\usepackage{graphicx}
\usepackage[misc]{ifsym}
\usepackage{comment} 
\usepackage{url}
\urldef{\mailsa}\path|iker.malaina@ehu.eus|

\begin{document}

\maketitle

\renewcommand{\abstractname}{English version of abstract}
\begin{abstract}
In this study, 529 variables extracted from dynamic contrast-enhanced magnetic resonance imaging (DCE-MRI) of 922 breast cancer patients have been evaluated, focusing on distinguishing between recurrent and non-recurrent cases, as well as those with and without metastasis. Special emphasis is placed on the differences among invasive breast cancer subtypes (Luminal A, Luminal B, HER2 positive, and TNBC). The accurate identification of the subtype is crucial as it impacts both treatment and prognosis. The analysis is based on the dataset from Saha et al., highlighting key factors for predicting recurrences and metastases, providing valuable information for proper monitoring and the selection of effective treatments.

\end{abstract}
 -----
\renewcommand{\abstractname}{Other language version of abstract}
\begin{abstract}
En este estudio, se han evaluado 529 variables extraídas de imágenes de resonancia magnética dinámica con contraste (DCE-MRI) de 922 pacientes con cáncer de mama, centrándose en discernir entre casos recurrentes y no recurrentes, así como entre aquellos con metástasis y sin ella, haciendo especial hincapié en las diferencias entre los subtipos de cáncer de mama invasivo (Luminal A, Luminal B, HER2 positivo y TNBC). La correcta identificación del subtipo es crucial, ya que afecta tanto al tratamiento como al pronóstico. El análisis se basa en la base de datos de Saha et al., y se destacan factores clave para prever recurrencias y metástasis, proporcionando información valiosa para un seguimiento adecuado y la elección de tratamientos eficaces.

\end{abstract}

\section{Introducción}
El cáncer de mama es una patología originada cuando las células mamarias comienzan a crecer de manera descontrolada, formando un tumor maligno en los tejidos mamarios. Este tipo de cáncer puede surgir tanto en mujeres como en hombres, aunque es significativamente más común en mujeres. Sus causas son multifactoriales, y pueden incluir rasgos genéticos, hormonales y ambientales. Además, mutaciones genéticas hereditarias, como las relacionadas con los genes BRCA1 y BRCA2, pueden aumentar el riesgo de desarrollar este cáncer \cite{genes}. Por otro lado, factores hormonales, como la exposición prolongada a hormonas sexuales femeninas, así como antecedentes familiares de cáncer de mama, pueden contribuir a esta susceptibilidad \cite{hormonas}.

La incidencia del cáncer de mama en España es significativa, proyectándose que en 2023 se diagnosticarán 35,001 nuevos casos, lo que lo convierte en el cáncer más común entre las mujeres, representando alrededor del 30\% de los cánceres femeninos diagnosticados. La tasa de incidencia estimada es de 132 casos por cada 100,000 habitantes, con una probabilidad de 1 de cada 8 mujeres de desarrollar cáncer de mama. Este cáncer tiende a manifestarse con mayor frecuencia entre los 45 y 65 años debido a los cambios hormonales en los periodos de peri y postmenopausia.

A pesar de los avances en detección temprana y tratamiento, el cáncer de mama sigue siendo la principal causa de muerte por cáncer en mujeres en España. Se proyecta que 6,528 mujeres fallecerán a causa de este cáncer en 2023, representando el 5.8\% de todas las muertes por cáncer en el país. La tasa de mortalidad es de 22.7 por cada 100,000 habitantes \cite{SEOM}. 

En cuanto a la recurrencia del cáncer de mama, hay estudios que indican que el 1.553\% de los casos sufrirán una reaparición del tumor en menos de 10 años\cite{recurrencia} llegando hasta un 15\% en algunos casos \cite{recurrencia2}, mientras que, por ejemplo, en Estados Unidos, cerca del 4\% de estos cánceres derivarán en metástasis \cite{metastasis} llegando, de acuerdo con algunos estudios, hasta el 21\% \cite{recurrencia2}, por lo que estas dos condiciones tienen una especial relevancia a la hora de ser predichas. 

La detección temprana del cáncer de mama es crucial para mejorar las tasas de supervivencia y los resultados del tratamiento. Las estrategias de detección suelen incluir mamografías regulares, la autoexploración mamaria y las revisiones médicas periódicas para identificar posibles cambios en los senos. Además, pruebas más avanzadas, como la resonancia magnética mamaria (MRI) y la biopsia, pueden utilizarse para evaluar lesiones sospechosas y confirmar la presencia de cáncer \cite{detec}

La MRI (Resonancia Magnética) es un método de diagnóstico por imagen que utiliza campos magnéticos y ondas de radio para producir imágenes detalladas del interior del cuerpo. Esta técnica no utiliza radiación ionizante, lo que la hace más segura y no invasiva. Su funcionamiento se basa en la resonancia magnética de los núcleos de hidrógeno presentes en el cuerpo, que, al ser expuestos a un potente campo magnético y pulsos de radiofrecuencia, generan información que se traduce en imágenes tridimensionales de los tejidos blandos. Esto posibilita una visualización minuciosa de órganos y estructuras internas.

La DCE-MRI (Resonancia Magnética Dinámica con Contraste) es una modalidad específica de MRI que utiliza la perfusión sanguínea y la distribución de contraste en los tejidos. Durante una DCE-MRI, se inyecta un agente de contraste en sangre para resaltar áreas específicas y obtener información dinámica sobre el flujo sanguíneo. En particular en el cáncer de mama, la DCE-MRI sirve para evaluar las características dinámicas de los tumores, como su vascularización, lo que es esencial para determinar la malignidad del tumor, planificar el tratamiento y evaluar la respuesta a la terapia \cite{dcemri}. 

En este trabajo, con el objeto de valorar la capacidad predictiva de la información extraída de las imágenes de resonancia magnética, por un lado, hemos estudiado 529 variables asociadas a características extraídas a partir de imágenes de DCE-MRI registradas durante más de una década en un mismo centro. Más concretamente, nos hemos centrado en identificar variables que puedan discernir entre los cánceres que dieron lugar a recurrencia posterior y los que no, y los que dieron lugar a metástasis y los que no. Por otro lado, hemos hecho hincapié en las diferencias de estas dos condiciones entre los cuatro principales subtipos de cáncer de mama invasivo (Luminal A, Luminal B, HER2 positivo y TNBC). La muestra utilizada está formada por la información de imágenes de DCE-MRI de 922 pacientes, cuyo diagnóstico fue confirmado posteriormente por biopsia. En particular, hemos analizado en nuestro estudio la base de datos \cite{orig1} disponible en el repositorio \cite{orig3}, utilizada por Saha et al. en \cite{orig2}.

La importancia de la correcta identificación de un subtipo de cáncer radica en que tanto el tratamiento como el pronóstico son diferentes dependiendo del tipo, y un abordaje apropiado y rápido puede marcar la diferencia en el desarrollo de la enfermedad, tanto en términos de recurrencia como en términos de metástasis \cite{tratamdif}.Por otro lado, una vez realizadas las pruebas de DCE-MRI apropiadas, identificar las variables adecuadas a la hora de predecir o diagnosticar estas condiciones puede ser de gran valor para el paciente, siendo la recurrencia o la metástasis dos condiciones muy nocivas que pueden suceder a futuro.

En definitiva, aquí, se presenta un análisis basado en imágenes de DCE-MRI donde se destacan los factores a valorar a la hora de poner el foco sobre en casos con mayor probabilidad de recurrencia o metástasis, lo que podría ayudar en el seguimiento y la elección del correcto tratamiento para combatir esta enfermedad de manera eficaz.

\section{Métodos}

Esta sección está dividida en las siguientes subsecciones:
\begin{enumerate}
    \item Información recogida en la base de datos. En este apartado se presentan las variables de la base de datos a partir de la cual se ha extraído el material analizado en este trabajo.
    \item Características de las imágenes. En esta sección se detalla el tipo de imágenes de resonancia magnética utilizadas para el diagnóstico del cáncer de mama.
    \item Subtipos de cáncer considerados. Aquí se explican los diferentes subtipos de cáncer que se van a analizar, caracterizándolos y mencionando su pronóstico.
    \item Software y análisis estadístico. En esta subsección se mencionan las herramientas y tests estadísticos utilizados para analizar las variables del estudio
   \end{enumerate}

\subsection{Información Recogida en la Base de Datos}

Como se indica en \cite{orig1}, además de las imágenes de MRI de los pacientes que se han utilizado en este trabajo, se almacenaron datos demográficos, características del tumor, seguimiento, etc. de los pacientes. En la Table \ref{tabla1} se resumen todas las variables recogidas en dicha base de datos.

\begin{table}
\centering
\caption{Información de los pacientes}
\label{tabla1}

\resizebox{15cm}{!} {
\hspace{-2cm}

\begin{tabular}{|l|l|}
\hline
\textbf{Categoría} & \textbf{Variables recogidas} \\
\hline
Demografía & \begin{tabular}[c]{@{}l@{}}
- Edad en la fecha del diagnóstico. \\
- Estado menopáusico al diagnóstico. \\
- Raza/etnia. \\
- Enfermedad metastásica en la presentación.
\end{tabular} \\
\hline
Características del Tumor & \begin{tabular}[c]{@{}l@{}}
- Estado del receptor de estrógeno. \\
- Estado del receptor de progesterona. \\
- Estado del receptor del factor de crecimiento epidérmico humano 2. \\
- Subtipo molecular. \\
- Oncotype score. \\
- Estadificación del cáncer. \\
- Grado del tumor. \\
- Grado de Nottingham. \\
- Tipo histológico. \\
- Ubicación del tumor. \\
- Posición del tumor. \\
- Cáncer de mama bilateral. \\
- Lado anotado en la imagen.
\end{tabular} \\
\hline
Hallazgos de RM & \begin{tabular}[c]{@{}l@{}}
- Multicéntrico/multifocal. \\
- Compromiso de la mama contralateral. \\
- Linfadenopatía o ganglios linfáticos sospechosos. \\
- Compromiso de la piel/pezón. \\
- Compromiso del músculo pectoral/tórax.
\end{tabular} \\
\hline
Cirugía & \begin{tabular}[c]{@{}l@{}}
- Estado de la cirugía. \\
- Días hasta la cirugía desde el diagnóstico. \\
- Tipo de cirugía definitiva.
\end{tabular} \\
\hline
Radioterapia & \begin{tabular}[c]{@{}l@{}}
- Radiación neoadyuvante. \\
- Radiación adyuvante.
\end{tabular} \\
\hline
Respuesta del Tumor & \begin{tabular}[c]{@{}l@{}}
- Respuesta clínica. \\
- Respuesta patológica a la terapia neoadyuvante.
\end{tabular} \\
\hline
Recurrencia & \begin{tabular}[c]{@{}l@{}}
- No, sí. \\
- Días hasta la recurrencia local y/o a distancia.
\end{tabular} \\
\hline
Seguimiento & \begin{tabular}[c]{@{}l@{}}
- Días hasta la muerte desde el diagnóstico. \\
- Días hasta la última evaluación libre de recurrencia local. \\
- Días hasta la última evaluación libre de recurrencia a distancia. \\
- Días hasta el último contacto.
\end{tabular} \\
\hline
Características de la mamografía & \begin{tabular}[c]{@{}l@{}}
- Edad en la mamografía. \\
- Densidad mamaria. \\
- Forma de la lesión. \\
- Margen de la lesión. \\
- Distorsión arquitectónica. \\
- Densidad de la lesión. \\
- Calcificaciones. \\
- Tamaño de la lesión.
\end{tabular} \\
\hline
Características de la ecografía & \begin{tabular}[c]{@{}l@{}}
- Forma, margen, tamaño y ecogenicidad de la lesión. \\
- Sólida. \\
- Sombras acústicas posteriores.
\end{tabular} \\
\hline
Datos de terapia & \begin{tabular}[c]{@{}l@{}}
- Quimioterapia. \\
- Terapia endocrina. \\
- Terapia Anti-Her2/Neu. \\
- Terapia neo-adyuvante. \\
- Respuesta patológica a la terapia neo-adyuvante. \\
- Respuesta casi-completa.
\end{tabular} \\
\hline
Información técnica de la resonancia magnética (MRI) & \begin{tabular}[c]{@{}l@{}}
- Días hasta la resonancia magnética desde el diagnóstico. \\
- Fabricante y nombre del modelo. \\
- Opción de escaneo. \\
- Potencia del campo. \\
- Posición del paciente y de la imagen del paciente. \\
- Volumen de bolo de contraste. \\
- Tiempo de repetición y tiempo de eco. \\
- Matriz de adquisición. \\
- Grosor del corte. \\
- Filas y columnas. \\
- Diámetro de reconstrucción, Campo de visión (cm) y Ángulo de inclinación.

\end{tabular} \\
\hline
\end{tabular}
}
\end{table}

\subsection{Características de las Imágenes}

Los elementos utilizados en este trabajo son imágenes de resonancia magnética (MRI) de cortes transversales de tejido mamario adquiridas mediante resonadores magnéticos con una fuerza de campo magnético de 1.5 o 3 Teslas en posiciones prona, es decir, acostada boca abajo, accesibles en \cite{orig1}. Además, se utilizaron las siguientes secuencias de MRI en formato DICOM (Digital Imaging and Communications in Medicine): 
\begin{enumerate}
    \item Una secuencia ponderada en T1 sin saturación de grasa (útiles para visualizar la anatomía y la estructura de los tejidos).
    \item Una secuencia ponderada en T1 con saturación de grasa antes del contraste (la saturación de grasa ayuda a mejorar la visibilidad de ciertas estructuras).
    \item Entre tres a cuatro secuencias después del contraste (para evaluar la captación de contraste y las características dinámicas de los tejidos).
\end{enumerate}

En esta cohorte, se utilizaron tres tipos de agentes de contraste: gadobutrol (Gadavist, Bayer Healthcare, Berlín, Alemania) para 2 pacientes (0.2\%), gadopentetato de dimeglumina (Magnevist, Bayer Healthcare, Berlín, Alemania) para 560 pacientes (60.8\%), y gadobenato de dimeglumina (Multihance, Bracco, Milán, Italia) para 263 pacientes (28.5\%). Por último, no se especificó el nombre del contraste para 97 pacientes (10.5\%).

A partir de dichas imágenes, en \cite{orig2} se realizó una revisión exhaustiva de la literatura sobre el procesamiento de imágenes de resonancia magnética (RM) de mama, el diagnóstico asistido por computadora y la radiómica para compilar un amplio conjunto de características como predictores, categorizándolas según su origen y el tipo de procesamiento de imágenes. En particular, se utilizaron subconjuntos específicos de secuencias de RM para diferentes tipos de características relacionadas con la mama, el tumor y el tejido mamario denso, y se describieron diversos enfoques de procesamiento de imágenes utilizados para extraer estas características. En \cite{listadometod} puede encontrarse una lista completa de estas características que se consideraron. En este trabajo, nos hemos basado en dichas características, que fueron obtenidas a partir de las MRI, para evaluar la capacidad de las técnicas de resonancia en detección. Todas ellas fueron analizadas en el estudio, pero solo se hace mención a las que sirvieron para caracterizar algún subtipo de cáncer, o se asociaron a recurrencia del tumor después del tratamiento, o a metástasis ).

\subsection{Subtipos de Cáncer Considerados}

En esta sección se resumen las propiedades principales de los diferentes subtipos que se consideraron en el estudio:
\begin{itemize}

\item Luminal A: Este subtipo es positivo para los receptores hormonales de estrógeno (ER) y/o progesterona (PR), lo que indica que las células cancerosas tienen receptores para estas hormonas. Además, presentan bajos niveles de proliferación celular (Ki-67), y es negativo para HER2. Suelen tener un pronóstico más favorable en comparación con otros subtipos.

\item Luminal B: También es positivo para los receptores hormonales (ER/PR), pero a diferencia de Luminal A, Luminal B exhibe niveles más altos de proliferación celular (Ki-67), y es positivo para HER2. Puede tener un pronóstico menos favorable que Luminal A y, en ocasiones, se asocia con una respuesta menos efectiva a la terapia hormonal.

\item HER2 positivo: Este subtipo es positivo para la sobreexpresión del Receptor-2 del factor de crecimiento epidérmico humano (HER2). HER2 es un gen que puede promover el crecimiento celular cuando está sobre expresado. Antes de los tratamientos específicos para HER2, este subtipo solía tener un pronóstico menos favorable, pero con terapias dirigidas a HER2, el pronóstico ha mejorado significativamente.

\item TNBC (Triple Negativo): Este subtipo carece de expresión de receptores hormonales (ER/PR) y de sobreexpresión de HER2. Se denomina "triple negativo" porque no responde a terapias dirigidas a estos receptores. Puede tener un pronóstico más desafiante debido a la falta de terapias hormonales o dirigidas a HER2. No obstante, la investigación está en curso para desarrollar estrategias de tratamiento más efectivas.

\end{itemize}

\subsection{Software y Análisis Estadístico}
Todos los análisis fueron llevados a cabo a través del software MATLAB, 2019b. Para comparar los resultados, primero se analizó la normalidad de los datos a través del test de Kolmogorov-Smirnov. Para las comparaciones múltiples, se utilizaron los tests ANOVA y Kruskal-Wallis, para el caso paramétrico y no paramétrico, respectivamente. Para las comparaciones bilaterales se utilizaron el t-test y el test de rangos de Wilcoxon, para el caso paramétrico y no paramétrico. Para estudiar si existe alguna relación no aleatoria entre variables categóricas, obtener los Odds ratio y sus intervalos de confianza (con un nivel de confianza del 95\%), se utilizó el test exacto de Fisher, que es más adecuado que el test $\chi^2$ cuando hay grupos con una $n$ pequeña. Para las comparativas se presentaron el p-valor, intervalos de confianza, y el estadístico correspondiente. 

\section{Resultados}
El objetivo principal de este trabajo es estudiar la recurrencia y metástasis del cáncer de mama. Para ello, primero, se depuró la base de datos mencionada en la sección anterior, lo que nos dejó con un total de 719 pacientes (excluimos los casos en los que no hubiera información sobre alguna de las variables de interés).

De ese conjunto, 470 fueron categorizados como subtipo Luminal A, 83 como subtipo Luminal B, 45 como HER2 positivo, y por último, 121 como TNBC. Como puede observarse, el tipo Luminal A fue muy mayoritario con respecto del resto, y sus porcentaje relativo  se ajustan a los presentes en la literatura \cite{porcentajes}. En cuanto a las recurrencias, hubo un total de 62, distribuidas entre los diferentes tipos de cáncer como: 34 del tipo Luminal A, 8 del tipo Luminal B, 6 del tipo HER2, y 14 del tipo TNBC. Finalmente, hubo 30 casos de metástasis confirmados, de entre los cuales: 16 fueron del tipo Luminal A, 6 del Luminal B, 5 del HER2, y por último, 3 del tipo TNBC.

A continuación, se estudiaron los Odds Ratio de los diferentes tipos de cáncer contra el resto, con el objeto de ver si alguno de ellos presentaba mayor recurrencia o metástasis, de acuerdo con los datos del estudio. Con respecto al subtipo Luminal A, los resultados fueron: Odds Ratio $=0.616$, p-valor $=0.071$ y $CI=(0.364, 1.041)$  para la recurrencia, y OR $=0.592$, p-valor $=0.172$ y $CI=(0.284, 1.233)$ para la metástasis; para el subtipo Luminal B se obtuvo: Odds Ratio $=1.15$, p-valor $=0.68$ y $CI=(0.527, 2.51)$  para la recurrencia, y OR $=1.843$, p-valor $=0.247$ y $CI=(0.732, 4.642)$ para la metástasis; ; en el caso de HER2, los resultados fueron: Odds Ratio $=1.698$, p-valor $=0.266$ y $CI=(0.689, 4.184)$  para la recurrencia, y OR $=3.245$, p-valor $=0.0336$ y $CI=(1.18, 8.926)$ para la metástasis; por último, para el TNBC se obtuvo: Odds Ratio $=1.499$, p-valor $=0.214$ y $CI=(0.798, 2.816)$  para la recurrencia, y OR $=0.538$, p-valor $=0.454$ y $CI=(0.161, 1.802)$ para la metástasis.

\section{Discusión}
En este trabajo, hemos presentado un análisis de los factores de riesgo asociados a metástasis y a recurrencia del cáncer de mama, valorando la capacidad de las imágenes de resonancia magnética para identificar dichas condiciones. Primero, hemos estudiado si existía mayor riesgo de padecer tanto metástasis como recurrencia del cáncer dependiendo del subtipo de cáncer (Luminal A, Luminal B, HER2 y TNBC), en nuestra cohorte de casi 1000 pacientes. Sorprendentemente, en contra de lo que indican algunos estudios \cite{metastasis}, no se encontró una mayor incidencia en ninguna de las condiciones para el subptipo Triple Negativo. Por otro lado, nuestros resultados indicaron que el tipo HER2 presentó mayor probabilidad de metástasis que el resto de subtipos, con Odds Ratio de 3.24, lo que puede interpretarse como que la metástasis de este subtipo tuvo una incidencia tres veces mayor que la del resto. 

Este trabajo aún es una versión preliminar del estudio. Para completar el análisis, trabajaremos con las variables obtenidas a partir de las MRI, buscando las más adecuadas para poder identificar tanto la recurrencia como la metástasis, y ayudar así al diagnóstico, la prevención, y el tratamiento de esta afección.

\end{document}